\input{aipcheck}

\documentclass[
    ,final            % use final for the camera ready runs
  ]
  {aipproc}

\layoutstyle{6x9}

\begin{document}

\title{The evolution of the cosmic SN rate}


\author{Enrico Cappellaro}{
  address={INAF, Osservatorio Astronomico di Padova, vicolo dell'Osservatorio 5, 
  35122 Padova, Italy}
}
\author{Maria Teresa Botticella}{
  address={INAF, Osservatorio Astronomico di Teramo}
}
\author{Laura Greggio}{
  address={INAF, Osservatorio Astronomico di Padova, vicolo dell'Osservatorio 5, 
  35122 Padova, Italy}
}

\begin{abstract}
We briefly review the contribution of SN rate measurements to the debate on SN progenitor scenarios. We find that core collapse rates confirms the rapid evolution of the star formation rate with redshift. After accounting for the dispersion of SN~Ia measurements and uncertainty of the star formation history, the standard scenarios for  SN~Ia progenitors appear consistent with all observational constraints.

 \end{abstract}

\maketitle

%%%%%%%%%%%%%%%%%%%%%%%%%%%%%%%%%%%%%%%%%%%%
%% MAINMATTER
%%%%%%%%%%%%%%%%%%%%%%%%%%%%%%%%%%%%%%%%%%%%

\section{Introduction}

The rate of supernovae (SNe) provides a link between the evolution  of individual stars and that of stellar systems. 

Stellar evolution theories predict that  all stars  more massive than 8-10~M$_\odot$  complete the eso-energetic nuclear burnings  to end with an iron core. The collapse of the iron core results in the formation of a compact object, a neutron star or possibly a black hole, accompanied by the high-velocity ejection of a large fraction of the star mass. SN~II derive from the core collapse of stars that, at the time of explosion, retain their H envelopes, whereas stars which lost their H (and He) envelope are thought to be the progenitors of SN~Ib (SN~Ic) \cite{Heger2003}. Given the short lifetime of their massive progenitors ($<30$~Myr), the rate of core collapse SN in a given stellar system directly traces the current star formation rate (SFR). Conversely, when the SFR is know, the rate can be used to verify the consistency of the progenitor scenario \cite{Hopkins2006}.

Stars of mass lower than 8 M$_\odot$ quietly terminate their life as white dwarfs (WDs) but for a small fraction which belong to special, close binary systems. In these cases,   mass transfer from the secondary star may take the WD above the Chandrasekhar limit of about 1.4~M$_\odot$. When this occurs the degenerate electron pressure no longer sustains the star, which is destroyed  by a thermonuclear explosion. These events are identified with SN~Ia. The elapsed time from  star formation to explosion (delay-time) depends not only on the evolutionary time of the progenitor star but also from that of the companion and from the orbital parameters. In general, the minimum delay-time is set by the evolutionary  lifetime  of the most massive companion of a WD ($\sim 30$~Myr) whereas, depending on the orbital parameters, it is possible to achieve delay-time longer than the present age of the Universe. While there is a wide agreement on this basic description, there are still many fundamental issues that are strongly debated, first of all the nature of the companion star, either a giant star filling the Roche lobe during its late evolution (single degenerate scenario, SD) or also a WD which merges with the primary WD after the orbital shrinking due to gravitational wave emission (double degenerate scenario, DD) \cite{Hillebrand2000}. The many uncertainties on initial conditions, details of the binary evolution and also on the actual explosion mechanism allow for a wide range in the predicted SN~Ia rate in different stellar systems\cite{Greggio2005}. Rate measurements as a function of redshift and/or properties of the stellar parent population can be used to restrict the viable scenarios, at least in principle.

\section{SN searches and SN rates}

The SNe discovered in the few  years of this century outnumber those discovered in the whole previous century \cite{cap05}. This is the result of many different
contributors ranging from the deep search for high redshift SN~Ia to be used as cosmological probes to  the thorough searches of nearby galaxies either by professional automated surveys or by an increasing number of very effective amateurs. In between,  new wide field CCD mosaic cameras at medium size telescope have boosted the search at medium redshift ($0.1<z<0.5$) where both large field of view and sensitivity are required.

A major fraction of the discoveries are SN~Ia (60\%), but this is an observational bias due to the selection criteria of high redshift searches. Indeed, considering only the volume up to redshift $z=0.01$ where we can assume that the searches are largely unbiased with respect to SN types, core collapse are 2/3 of all SNe. Of the core collapses,  3/4 are type II and the others type Ib/c (cf. the Asiago Supernova Catalogue, http://web.oapd.inaf.it/supern/snean.txt). 

The betters statistics and handling of systematics is being exploited to obtain more accurate estimates of the SN rates. In particular for what concern the rate in the local Universe (cf. Weidong Li, this conference) the goal is to update the current best estimates still based on past photographic and visual searches \cite{stat99}. Waiting for this work to be completed,  the existing SN list  can be used to get some directions on current estimates. To this aim we selected from the RC3 catalogue \cite{RC3} all the galaxies with redshift $z<0.01$ and, based on the SN rates published by \citet{stat99}, we predicted the number of SNe of the different types which are expected out of this sample of nearby galaxies of known morphological types and luminosities. These expected numbers were compared with the actual discovery statistics of the last 5 yr (2001-2005):  while for SN~Ia and SN~II the expected and observed numbers  appear consistent, for SN~Ib/c the estimated number is significantly lower than the actually detected events  which suggests that the rate for these SNe in \citet{stat99} is underestimated by a factor $\sim 2$.  Possibly this is due to misclassification of  SN~Ib/c in the photographic sample since this class of SNe  was recognized only in the mid '80 when most of the photographic searches were close to their end.

The new data confirm the dependency of SN rate on galaxy types with SN~Ia showing a constant rate (per unit luminosity) from elliptical to spiral. The rates of SN~II and SN~Ib/c, instead,  peak in late spirals, similar to the SFR \cite{Kennicutt} which establishes a direct connection between core collapse SNe and massive stars. On the other hand, the different behavior of SN~Ia seems to call for apparently conflicting requirements: the fact that SN~Ia are found so numerous in Elliptical, where star formation is very low (if any) suggests long-lived, hence low mass, progenitors \cite{VDB1959,Bertaud1961}. However, since a major fraction of the blue luminosity in spirals is due to massive stars, at a given luminosity spirals host a lower number of low mass systems than ellipticals. This means that in spirals a major fraction of SN~Ia have young progenitors \cite{oemler_tinsley}. The apparent contradiction was solved in  the early '80 showing that binary evolution allows for a wide range of delay times with a distribution which favours the short delay times \cite{Greggio1983}. 

Recently, this issue was revived by new measurements of the SN rates with redshift. 
In most cases these measurements are a by-product of high redshift searches for SN~Ia  which explains why there is a far better sampling for this SN type than for the core collapse. The SN~Ia rate appears to show a rapid increase up to redshift $\sim 1$ and a turn-down at higher redshift. It was claimed that this behavior requires for all SN~Ia a long  delay time of $\sim 3-4$ Gyr \cite{strolger04,strolger06}. This results is still to be confirmed due to poor statistics at high redshift, but also in view of the large dispersion of different measurements at lower redshift (see next section). 
Taken at face value however, this  conflicts with the new evidences found in the local Universe, where the rate per unit mass as a function of galaxy color \cite{Mannucci2005} requires that a significant fraction of progenitors are young (a similar
claim based on the history of metal enrichment \cite{Scannapieco} appears disputable).

Actually, the observed, high rate of SN~Ia in radio-loud ellipticals  has been interpreted as evidence for very short delay times ($<100$~Myr) for  a large fraction ($\sim 50$\%) of SN~Ia progenitors \cite{DellaValle2005} under the hypothesis that both the radio emission and the enhanced SN~Ia rate are due to recent bursts of star formation, supposed to occur at random in the lifetime of all ellipticals following merging episodes. We have to note that in this case we would expect also a number of core collapse SNe to be found in E. To date, only 3 core collapses (5\% of all SNe) have been found in Ellipticals  compared with 58 SN~Ia. However, in star forming galaxies, such as late spirals, we detect 1 SN~Ia every 2.5 core collapse events. If the conditions are the same in radio loud ellipticals we have to conclude that only 1 SN~Ia out of the 58 observed is related to recent SF.  Either the present low level SF in ellipticals produces stars with a IMF strongly biased toward low mass stars, or we have to search for an alternative explanation. Before that, it is mandatory to verify the high rate of SNIa in radio loud ellipticals using a larger sample.

We conclude that at present the data for the local Universe are fully consistent with the prediction of the standard scenarios of stellar evolution but for the above mentioned peculiarity of radio loud ellipticals.

\subsection{STRESS: the Southern InTermediate Redshift ESO SN Search}

STRESS is the SN search we have carried on at ESO using the WFI at the 2.2m telescope for candidate detection, complemented with FORS at the VLT for spectroscopic confirmation. Different from other high redshift searches, STRESS was especially designed  to measure SN rates and to reduce the candidate selection biases with respect to the different SN types. In particular we did not exclude candidates found close to the galaxy nucleus, which explains the large contamination of our candidate list with variable AGNs.

Preliminary analysis showed a very rapid increase of the core collapse rate with redshift  which appeared consistent with the more recent estimate of SFR evolution \cite{stat05}. We have now completed the analysis of all our data which will be described in detail in a forthcoming  paper \cite{Botticella} and briefly sumarized here.

The search produced about 200 SN candidates, 60 of which turned out to be AGNs based on the long term variability history. We obtained direct spectroscopy for  41 candidates: 15 were SN~Ia, 19 core collapse (of which 5 type Ib/c) and 7 AGN. For other 44 candidates  we also obtained spectra of  the host galaxy, 22 of which were AGN.
Multi-color observations were used to characterize the galaxy sample, to derive photometric redshifts, absolute luminosities and rest frame colors for all galaxies down to R=21.8 limit. The final SN list, including only events occurring in one of the galaxy of the sample, is made of 26 spectroscopically confirmed SNe, 20 SN candidates with host galaxy spectroscopy, and 44 candidates with only photometric redshift.  The remaining 31 which were found close to the host galaxy nucleus were given a 0.5 weight in the statistics. 

Particular care was devoted to estimate statistical and systematic errors and to the accurate modeling of the extinction correction, one of the most uncertain steps in the computation. It turns out that despite our relatively low statistics, systematic errors dominate with two main contributors: the lack of spectroscopic confirmation for all candidates and the uncertainty in the extinction correction.

As  a result of this effort we obtained a measure of the SN rate and rate evolution in the redshift range covered by our search, that is $0.05<z<0.6$. The measured rate at the mean search redshift is shown in Tab.~\ref{tabres}. To simplify the comparison with the local estimates, we normalized the rate to the galaxy blue luminosity; then, for a  comparison with other measurements in the literature, we converted estimates in rate per unit volume multiplying by the proper luminosity density \cite{Botticella}.

The SN~Ia rate in SNu appears to be almost constant up to $z=0.3$, whereas the SN~CC rate grows by a factor 2 already at $z=0.21$.
This implies that the $r^{CC}/r^{Ia}$ ratio increases of a factor $\sim 2$ from  the local Universe to a look-back time of "only" 3 Gyr. 
If we consider that for the same look-back time the cosmic SFR increases by a similar factor ($2-3$), the evolution with redshift of the ratio $r^{CC}/r^{Ia}$ requires that a significant fraction of SN~Ia progenitors has lifetime longer that $2-3$ Gyr.

\begin{table}[!t]
\begin{tabular}{lcccp{.7\textwidth}}
\hline
\tablehead{1}{r}{b}{SN type}     &
\tablehead{1}{c}{b}{$\overline{z}$}& 
\tablehead{2}{c}{b}{SN rate}\\
 & &  
\tablehead{1}{c}{b}{ [SNu $h^2$]}   &
\tablehead{1}{c}{b}{$10^{-4} \mbox{yr}^{-1} \mbox{Mpc}^{-3}\, h^3$]}  \\
\hline\\
\medskip
SNIa         & $0.30^{+0.14}_{-0.14}  $    &   $0.22^{+0.10 +0.16}_{-0.08 -0.16}$ 
                                  &      $0.34^{+0.15 +0.25}_{-0.12 -0.15}$  \\
\medskip
SNCC      & $0.21^{+0.08}_{-0.09}$     &  $0.82^{+0.31 +0.30}_{-0.24 -0.26}$  
                              &     $1.14^{+0.43 +0.41}_{-0.33 -0.36}$  \\                   
\hline
\end{tabular}
\caption{SN rate measurements from STRESS ($h=H/70$).}\label{tabres}
\end{table}

\subsection{Core collapse rates}

\begin{figure}
\includegraphics[height=.47\textheight]{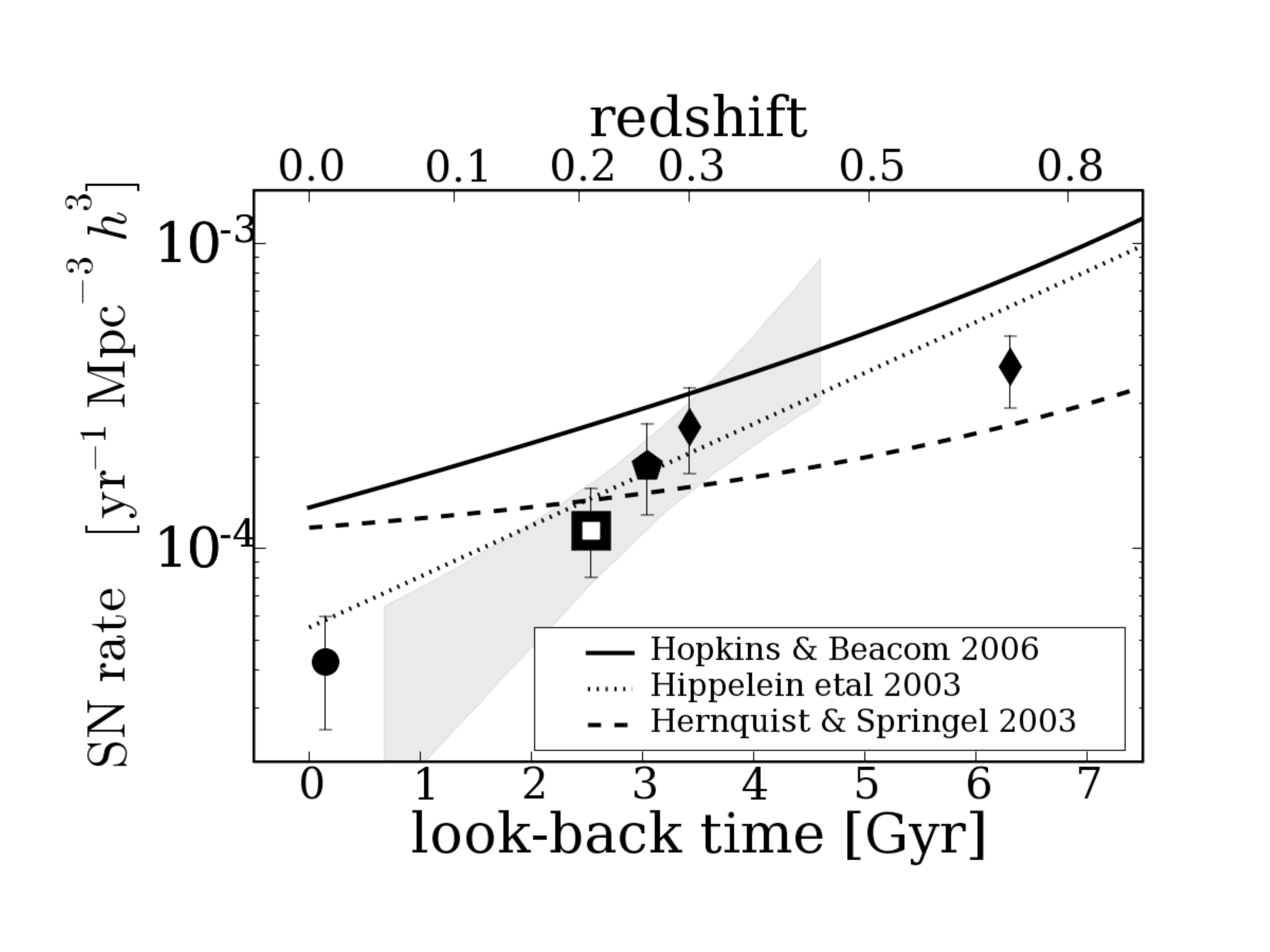}

  \caption{Core Collapse rates measurements with look-back time (legend as in Fig.~3). The shaded area is the 1-$\sigma$ confidence level of the rate evolution derive with STRESS. The lines are the predicted SN rate based on different SFR evolution from literature with $K^{CC}=0.09$ (see text). }\label{ccrate}
\end{figure}

There is a simple direct relation between the SN~CC and SF rate, namely $r^{\rm CC}(z) = K^{\rm CC} \times \psi(z) $, where $\psi(z)$ is the SFR and $K^{CC}$ is the number of stars per unit mass which end up as core collapse. For a SalA IMF and a standard $8-50$ $M_{\odot}$ range for SN~CC progenitors,  $K^{\rm CC}=0.009$. 
A few representative SFR history are shown in Fig~\ref{ccrate} after conversion for the above   $K^{\rm CC}$ factor. The rate measurements agree with the steep increase with redshift of recent estimate of the  SFR \cite{Hopkins2006,Hippelein}.
 For a look-back time of 3~Gyr ($z=0.25$) both the SFR and the SN~CC rate increase by a factor $\sim3$ compared to the local value. Instead, flatter SFR evolution \cite{Hernquist}  appears inconsistent with the observed SN~CC rate.
 
Aside from the rate evolution, the actual values of the SN~CC rates appear in excellent agreement with the SFR measured from  $H\alpha$ luminosity, while SFR derived from FIR observations imply, in general, significantly higher rate \cite{Hopkins2006}. 
On the one hand we may think that the extinction correction for SN~CC was underestimated. Alternatively, we may squeeze the mass range for SN~CC progenitors. In particular taking the lower limit to $10-12$ $M_{\odot}$ will solve the issue. This seems at odd with some recent direct measurements of SN~CC progenitor masses \citep{vandyk,Smartt,Li2006}, although it may be consistent with stellar evolution theory if high mass loss operates during AGV evolution of low mass stars, leaving a ONeMG WD remnant \cite{ritossa}.

\subsection{SNIa progenitor population}

For the SN~Ia rate, the relation with SFR is mediated by the distribution of the delay times:

\begin{figure}
\includegraphics[height=.35\textheight]{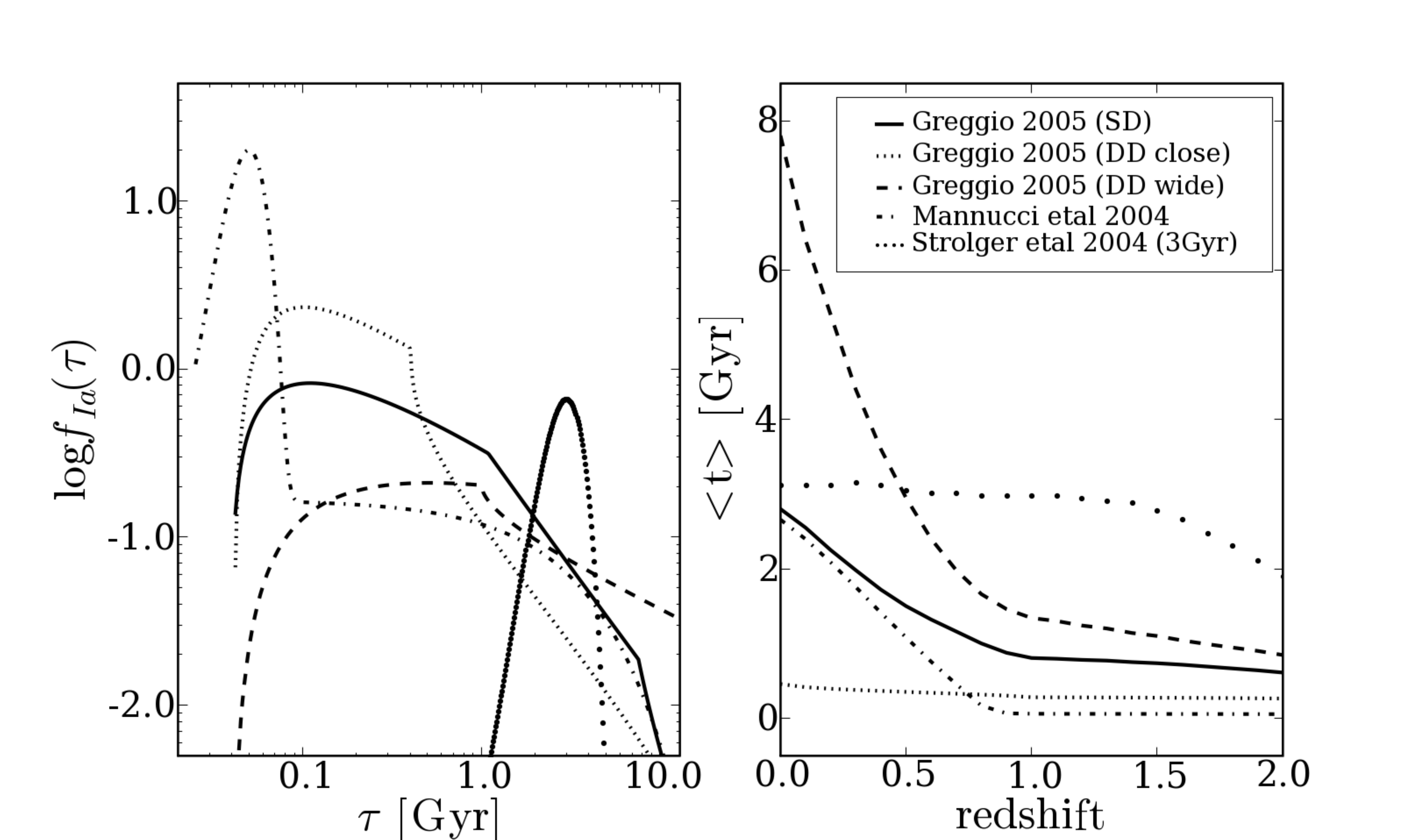}
  \caption{Left panel: delay time distribution for different models of SN~Ia progenitors.
  Right panel: median age of SN~Ia progenitors as a function of redshift. The latter was derived from the delay time distributions adopting the cosmic star formation history
from \citet{Hopkins2006}. }\label{dtd}
\end{figure}

$$\label{sniaeq}
r^{\rm Ia}(t) = k_{\alpha} A^{\rm Ia} \int_{\tau_{i}}^{min(t,\tau_{x})} f^{\rm Ia}(\tau) \psi(t- \tau)d{\tau}
$$

where $k_{\alpha}$ is the number of stars per unit mass of the stellar generation born at epoch $t-\tau$, $A^{\rm Ia}$ is the realization probability of the SN~Ia scenario , $f^{\rm Ia}(\tau)$ is the distribution function of the delay times (DTD), $\psi(t-\tau)$ is the SFR at epoch $t-\tau$, $\tau_i$ and $\tau_x$ are the minimum and maximum delay times for a given progenitor scenarios \cite[for details see][]{Greggio2005}.

\begin{figure}
\includegraphics[height=.47\textheight]{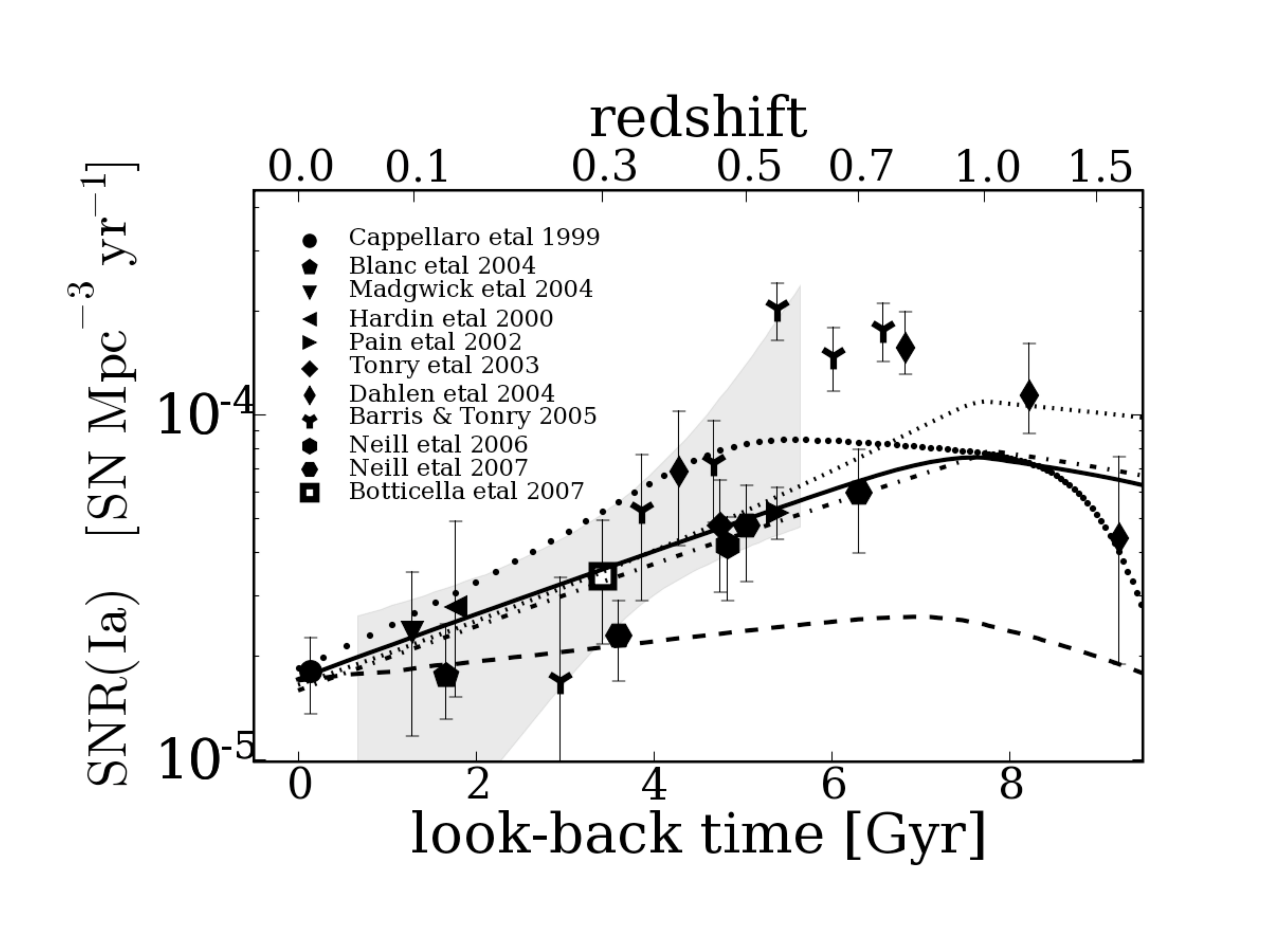}
  \caption{SN~Ia rate evolution with look-back time (the shaded area is the 1-$\sigma$ confidence level of the rate evolution derive with STRESS) .  Lines are the predicted SN rate assuming the DTD shown in the left panel of Fig.~2 and the cosmic star formation history from \citet{Hopkins2006}.}\label{ria}
\end{figure}

Recently, it has been argued that the SN~Ia rate evolution at high redshift is better fitted by a DTD with a gaussian shape centered at 2-4 Gyr \cite{strolger06} (see Fig.~2). 
Actually,  given the uncertainties on both SFH and SN rate measurements, these constraints on the progenitor model are week \cite{Forster}.
In any case, as we have mentioned in the previous section, different observations suggest a wide range for the delay times of SN~Ia progenitors. This leads other authors to propose an empirical DTD with two contributions, a "prompt" component proportional to the on-going SFR,  and a "tardy" component, described by an exponential function with a decay time of about 3 Gyr \cite{Mannucci2006}. 

A different approach is to derive the DTD from a detailed analysis of the astrophysical scenarios for the evolution of the binary systems which are candidate to produce SN~Ia. This approach has the advantage that the SN rate observations can be used to rule out some of the candidate systems, at least in principle. Analytical formulations of the DTD functions for different scenarios of SN~Ia progenitor have been derived by \cite{Greggio2005}   and have been used to predict the evolution of the SN~Ia rate for different  SFH \cite{blanc}. Here we limit the discussion to three representative cases, a standard SD model and the two DD models which produce the more extreme DTD, namely a "close DD"  and a "wide DD"  (Fig.~\ref{dtd}).

The predicted evolution of the SN~Ia rate for different DTDs is compared to the actual measurements in Fig.~\ref{ria}, where the value of $k_\alpha \, A^{Ia}$ was fixed to match the local rate. 

As it can be seen from the figure there is a large dispersion of different measurements, in particular at medium redshift,  $0.4<z<0.7$ which would deserve more attention. Our measurement seems to sits in between, though, given the large errors, it is actually consistent with all other observations. Given this large dispersion none of the considered DTD can be ruled out, with the exception of the "wide" DD model which predicts a too flat evolution.

Yet, with the adopted SFH, none of the DTD considered here is able to fit the very rapid increases of the SN~Ia rate up to z=0.5 which is suggested by some observations, with the possible exception of the DTD of \citet{strolger04} which also fits, by construction,  the rate decline at redshift $z>1$. We have to stress however that this DTD fails to reproduce the dependence of SN rate on galaxy colors which is observed in the local Universe \cite{Mannucci2005}.  A distinctive feature of this DTD is that the average age of SN~Ia progenitors at the time of explosion is 3 Gyr, at all redshift up to $z\simeq1.5$  (Fig.~2). For all other DTDs, high redshift SN~Ia progenitors are young.

Clearly there is still work to be done to reduce the systematic of  SN rate measurements. Meanwhile there seems no compelling motivations to deviate from standard stellar evolution. 

%%%%%%%%%%%%%%%%%%%%%%%%%%%%%%%%%%%%%%%%%%%%
%% SAMPLE TABLE
%%
%% Shows the use of \tablehead and \tablenote
%% macros
%%%%%%%%%%%%%%%%%%%%%%%%%%%%%%%%%%%%%%%%%%%%

%%%%%%%%%%%%%%%%%%%%%%%%%%%%%%%%%%%%%%%%%%%%%%%%
%% BACKMATTER
%%%%%%%%%%%%%%%%%%%%%%%%%%%%%%%%%%%%%%%%%%%%%%%%

\begin{theacknowledgments}
This research was funded by the program PRIN-MIUR 2004. 
\end{theacknowledgments}



\bibliographystyle{aipproc}   % if natbib is available

\bibliography{cappellaro}

\IfFileExists{\jobname.bbl}{}
 {\typeout{}
  \typeout{******************************************}
  \typeout{** Please run "bibtex \jobname" to optain}
  \typeout{** the bibliography and then re-run LaTeX}
  \typeout{** twice to fix the references!}
  \typeout{******************************************}
  \typeout{}
 }

\end{document}